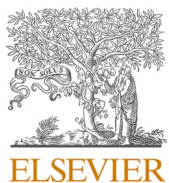
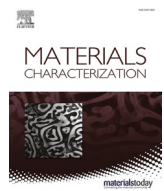

# Exploring nanoscale metallic multilayer Ta/Cu films: Structure and some insights on deformation and strengthening mechanisms

Daniel Karpinski [a], Tomas Polcar [a, *], Andrey Bondarev [a, b, *]

[a] Department of Control Engineering, Faculty of Electrical Engineering, Czech Technical University in Prague, Technicka 2, Prague 6 16627, Czech Republic
[b] Bernal Institute, School of Engineering, University of Limerick, V94 T9PX Limerick, Ireland

## ARTICLE INFO

## ABSTRACT

Nanoscale metallic multilayer (NMM) films are systems offering insight into the role of interfaces in metal plasticity, deformation, and strengthening mechanisms. Magnetron sputtering was used to fabricate the Ta/Cu NMM films with a periodicity (equal Ta and Cu layer thickness) from 6 to 80 nm, and with the structure exhibiting immiscible tetragonal β-Ta and face-centred cubic Cu phases. Transmission electron microscopy and X-ray diffraction analyses revealed that, irrespective of the period, all films manifested a polycrystalline structure. The growth direction of both Cu and Ta layers was found to be along $\langle 001 \rangle_{\beta\text{-Ta}} \parallel \langle 111 \rangle_{Cu}$ directions, with the crystallite size constrained by the layer thickness. The studies showed that the Ta/Cu NMMs exhibited compressive residual macro-stress and flow strength, and enhanced elastic recovery at the periodicity of ≤12 nm. Activation volume $V^*$ value of 11 to 20 $b^3$ as determined from the indentation creep test under a constant load, may indicate a mixed deformation mechanism. This mechanism likely involves the emission of dislocations from the incoherent Ta/Cu interfaces, as well as the formation of screw dislocations within Cu grains. The high-load indentation test, TEM studies, and the rCLS model collectively demonstrate that all NMM films predominantly undergo plastic deformation. This plastic deformation primarily occurs within the soft Cu layer, while the propagation of dislocations across the incoherent interface is largely excluded.

## 1. Introduction

The design of thin films with high strength is crucial for numerous industrial applications as protective layers against various types of wear. Typically, as the hardness of the film increases, its plasticity decreases, making it more brittle, less damage-tolerant, and prone to cracking [1]. On the contrary, a ductile film is less prone to cracking, but its softness increases susceptibility to wear. Therefore, the most crucial parameter is high film toughness, as it combines high hardness (strength) and high ductility. Achieving high toughness in thin films presents a challenging and demanding task. Researchers utilize intricate designs of film structures and microstructures to create composites, with the aim of enhancing toughness. Typically, toughness arises in composite films when a ductile component is incorporated into the hard and brittle matrix. An alternative method for controlling film toughness is through the fabrication of nanoscale multilayers using the physical vapor deposition technique [2–4]. Here, the toughness is adjusted by controlling the number and type of the interface between the elastically and plastically deformed layers, achieved through variations in bilayer thickness ($\Lambda = \lambda_{Mat1} + \lambda_{Mat2}$; where $\lambda$ is the individual layer thickness). When the $\lambda$ decreases from roughly 10 nm to several nm, so called superlattice structures in the crystalline multilayer system can be formed. These multilayer films are characterized by exceptional material properties, including higher hardness. In the superlattice structures, the individual layers are so thin that the material hardness is enhanced due to: (i) restricted dislocation movement in grains, (ii) high strain in an incoherent interface (high lattice mismatch on the interface of layers), and (iii) Koehler's stress [5,6]. Tantalum-copper (Ta/Cu) nanoscale metallic multilayers (NMMs) represent a system with a considerable difference between the mechanical properties of individual layers. Tantalum is a very hard metal; in a film form, it usually appears in two phases depending on the deposition parameters: the thermodynamically stable α-Ta with a body-centered cubic (*bcc*) structure or the metastable β-Ta with tetragonal (*t*) structure, which significantly differ in their mechanical properties [7,8]. The α-Ta phase is ductile with hardness ranging from 5 to 12 GPa, whereas the β-Ta phase is brittle, exhibiting higher hardness in the range of 15 to 19 GPa [7,8]. These remarkable properties of Ta are due to its more covalent character of the

---

metallic–covalent bonds caused by an almost half-filled d orbital ([Xe] 4 $f^{14}$ $5d^3$ $6s^2$). While Cu has a face-centred cubic (*fcc*) structure, it is a soft and ductile metal with a low hardness of 1–2 GPa, due to a fully filled d orbital ([Ar] $3d^{10}$ $4s^1$) which allows only metal bonds by half-filled s orbital, Cu exhibits a smaller overlap of the d-orbitals than Ta.

The mechanical properties of materials are directly influenced by their deformation mechanisms. Therefore, understanding these mechanisms is crucial for designing tailored NMMs with exceptional mechanical properties. In materials featuring a high density of interfaces, integrating interface physics like nucleation, motion, and reactions into existing models remains a challenge. This challenge arises partly from a limited understanding of deformation mechanisms primarily governed by interfaces, owing to the intricate nature of dislocation-interface interactions. The Ta/Cu NMMs were chosen for their unique combination of a very hard and brittle Ta layer with a hardness exceeding 15 GPa, along with a soft and ductile Cu layer, with a hardness of approximately 1.5 GPa. The β-Ta and *fcc* Cu multilayer films are expected to form incoherent interfaces, due to high mismatch of their lattice parameter, which could play an important role in hardness enhancement, according to the literature [9]. Previous research works on multilayered Ta/Cu materials have explored various aspects including: mechanical properties [10], indentation-induced shear band formation [11], plastic deformation and creep behavior [12,13], radiation tolerance of Ta/Cu amorphous/crystalline structure under He ion irradiation [14], strain rate sensitivity with different Ta/Cu modulation [15], and the effect of interface amorphization on mechanical properties of Ta/Cu NMMs [16]. In addition, efforts have been made to investigate the deformation behavior and annealing behavior of amorphous/crystalline CuTa/Cu NMMs and constant Cu layer thickness and different Ta content in CuTa layer [17–19]. This study explores the structural and mechanical characteristics, as well as the mechanisms governing strengthening and deformation, within nanoscale metallic Ta/Cu multilayered systems fabricated by PVD technique. Until now, the interrelationship between periodicity, microstructure, strengthening and deformation mechanisms of the crystalline Ta/Cu NMM films remains unknown. In this work, we provide a comprehensive structural analysis via transmission electron microscopy in combination with micro and nanomechanical measurements to correlate fine structure with observed mechanical behavior. Furthermore, the current work represents an initial endeavor to apply the previously established refined confined layer slip model to materials featuring a specific *fcc/tetragonal* copper/tantalum interface structure.

## 2. Experimental details

### 2.1. Film synthesis

The Ta/Cu nanoscale metallic multilayered films (1.90–2.15 μm thick) were deposited by magnetron sputtering using an AJA International Orion 4 system which was evacuated by a turbomolecular pump supported by multi-stage roots pump. The base pressure before each deposition was $2 \times 10^{-4}$ Pa. The chamber is equipped with four unbalanced magnetrons with mirror-like magnetic configuration between the individuals – the plasma was diverted to the chamber walls. Two magnetrons were equipped with circular indirectly cooled targets (⌀50.8 mm, and 6.35 mm thick): Cu (99.99% purity) and Ta (99.95% purity). Both were independently sputtered using direct-current (dc) power supplies. All films were deposited onto polished and ultrasonically pre-cleaned single-crystalline Si(100) substrates. Prior to the deposition, the substrates were etched by RF plasma at 50 W, and low argon pressure ($p_{Ar}$) of 0.4 Pa for 15 min. To reduce the effect of the crystalline substrate on growing NMM films, an amorphous 50 nm thin $Si_3N_4$ seed layer was deposited onto Si prior to Ta/Cu sputtering. The depositions of the Ta/Cu NMM films were conducted in argon gas at a low $p_{Ar} = 0.3$ Pa, RF substrate bias of −50 V, and without any external heating (a substrate temperature didn't exceed 50 °C). The substrates were rotated below the targets at a speed of 10 rpm, which were located at a target-to-substrate distance of ~13 cm. Two series of the Ta/Cu NMMs with low and high (approx. 2 times higher) deposition rate ($a_D$) was deposited. The dc power ($P_t$) applied on the Cu and Ta target in the case of a low $a_D$ was $P_{t\,Cu} = 100$ W and $P_{t\,Ta} = 200$ W, and in the case of a high $a_D$ was $P_{t\,Cu} = 200$ W and $P_{t\,Ta} = 350$ W, respectively. The bottom and top layer was always Ta.

### 2.2. Structural and chemical characterization

The total film thickness ($h_f$), and their curvature were measured using a 3D optical profilometer (NewView 7200, Zygo). The average biaxial residual macro-stress ($\sigma$) was calculated from Si substrate curvature by using Stoney formula. The crystalline structure of the NMMs was characterized using an X-ray diffraction spectrometer (Smartlab 9 kW, Rigaku) in parallel-beam configuration, using Cu$_{K\alpha}$ radiation (wavelength of 0.154 nm), with D/teX Ultra 250 silicon strip detector, and using omega-offset of 5° (to suppress Si(100) peak). The cross-section and indentation imprint images were obtained by a high-resolution scanning electron microscope (SEM) (Verios 460 L, Thermo Fisher Scientific). The cross-section lamellas of selected NMMs were prepared by a focused ion beam (FIB) milling with Ga$^+$ ions in the operating range of 2–30 kV using an SEM/FIB instrument (Helios NanoLab 660, Thermo Fisher Scientific). The selected area electron diffraction (SAED) patterns, nanobeam electron diffraction (NBD) patterns, high-angle annular dark-field (HAADF), bright-field (BF) images, and elemental composition maps of the films were acquired using transmission electron microscopes: Tecnai TF20 X-Twin and mono-chromated and image-corrected Titan Themis$^3$ (Thermo Fisher Scientific) equipped with energy-dispersive X-ray windowless Super-XG1 detector. Electron diffraction data were processed using CrysTBox software [20].

### 2.3. Mechanical characterization

The hardness ($H$), and effective Young's modulus ($E^*$) of the NMMs were evaluated from the load vs. displacement curves using Oliver and Pharr method [21]. Both mechanical properties were measured at room temperature and in an ambient environment by a nanoindenter (TI 950 Triboindenter, Hysitron) equipped with a Berkovich-type diamond tip at a load of 4 mN. The indentation depth was below 10% of film thickness, to avoid the substrate influence. The deformation behavior of the NMMs was tested by using an indentation test at a high indentation load of 200 mN, where the depth was around 50% of film thickness. Indentation creep tests were conducted with a constant load of 4 mN, and loading and unloading time was 10 s. The depth at a constant load was held and recorded for 100 s. The strain rate sensitivity ($m$) was determined from the linear part of the slope of the double-logarithmic plot of $H$ (flow stress) and imposed nominal strain rate [15,22] under isothermal conditions. The activation volume ($V^*$) was determined by the equation shown and described in [22]. The first estimation of a contact area (for Berkovich indenter) for average stress (hardness) calculation was determined according to the equation described in [23]. The flow strength ($\sigma_f = H/\alpha$) was estimated as hardness divided by Tabor factor α = 2.7 (for conventional metals, it is usually between 2.7 and 3.0 [24,25]).

## 3. Results and discussion

The Ta/Cu nanoscale metallic multilayered films were deposited with periodicity (bilayer thickness) Λ of 6, 12, 20, 40 and 80 nm, where the thickness of the Ta and Cu layer $\lambda_{Ta}$ and $\lambda_{Cu}$ was kept similar, i.e., $\lambda_{Ta} \approx \lambda_{Cu}$, respectively. The films were deposited at a low $p_{Ar} = 0.3$ Pa, resulting in their high energetic bombardment by fast neutral atoms (Ta and Cu), with energy up to several tens of eV [26,27], and almost without collisions of high-weight Ta (181 amu) atoms with working gas atoms (Ar). Particularly, the high energy of fast neutrals (Ta) contributes to a formation of compressive macro-stress during the deposition of the Ta/Cu NMMs [27,28]. Moreover, the backscattered and neutralized argon ions from the Ta target toward the substrate (with an energy of 380 eV and 450 eV, in the case of low and high $a_D$, respectively) also contributed to compressive macro-stress development in the films.

### 3.1. Structural properties

The XRD patterns of the as-deposited Ta/Cu NMMs with different $\Lambda$ deposited at a low and high $a_D$ as well as reference samples of pure Cu and Ta films deposited at a low and high $a_D$ are shown in Fig. 1a,b, respectively. The XRD pattern of the reference Ta film reveals the polycrystalline structure of the thermodynamically metastable tetragonal β-Ta phase (space group P 4‾ $2_1$ m) with (002) preferential orientation. The reference Cu film exhibits a polycrystalline structure of the *fcc*-Cu phase (space group F m 3‾ m) with a strong preferential orientation of (111), also signal from (220) plane is detected. The Ta/Cu NMMs with $\Lambda \geq 12$ nm exhibit two strong textures of the β-Ta(002) phase and *fcc*-Cu(111) phase with contribution of the broad β-Ta (004) peak. At $\Lambda = 6$ nm, the Ta/Cu NMMs exhibit low-intensity broad peaks of both phases. The satellite peaks observed around the main high-intensity peaks of $\Lambda = 12–20$ nm NMMs are caused by superlattice structure, like Zr/Nb NMMs [22,29]. Besides that, all NMMs exhibit low-intensity broad peak observable at $2\Theta = 38°$, suggesting the presence of a small amount either of an amorphous (a-) Ta—Cu phase at the Ta—Cu interface (see HRTEM analysis of the Ta—Cu interface), or α-Ta phase, or presence of both. The formation of the amorphous Ta—Cu interface, indicated by a broad peak at ~38°, was observed elsewhere [16]. H. Gong et al. revealed that ion-beam mixing [30], and annealing-driven interdiffusion between two constituent Ta and Cu layers at 600 °C [31] causes interfacial amorphization of the Ta/Cu interface. The Ta/Cu NMMs were exposed to high energy fast neutral Ta atoms during a low Ar pressure deposition, which can cause Ta—Cu intermixing at the Ta/ Cu interface. The presence of thermodynamically stable α-Ta(110) phase in the Ta/Cu NMMs can be ruled out due to a low deposition temperature. The α-Ta is usually expected at deposition temperatures higher than that of 300 °C [32]. Formation of the metastable β-Ta phase is facilitated by the non-equilibrium deposition process during the magnetron sputtering and should be stable at substrate temperatures up to 400 K [33]. Moreover, a small amount of oxygen from the residual atmosphere during the deposition facilitates the formation of the metastable β-Ta phase [7,34]. In our case, there is a very small amount of oxygen as observed by EDX, even when the base pressure before the deposition was low $2 \times 10^{-4}$ Pa; more details are shown in the next section.

Table 1 summarizes structural parameters extracted from the patterns shown in Fig. 1. The crystallite size $D$ (size of diffraction domains) for individual layers of the NMMs structure was roughly estimated by using the Scherrer equation. As expected, the $D$ proportionally decreased with $\lambda$. For $\lambda \leq 10$ nm, the $D$ is almost equal to $\lambda$, while for $\lambda > 10$ nm, the $D$ is almost two times lower than that of $\lambda$. No correlation between the $a_D$ and the $D$ was found at the nanoscale level, while thick Ta and Cu films exhibit a decrease in the $D$ with an increase of the $a_D$. In the case of *d*-spacing for Ta(002) ($d_{Ta} = 0.265$ nm) and Cu(111) ($d_{Cu} = 0.209$ nm), on the one hand, with decreasing $\Lambda$ below 12 nm, the *fcc*-Cu (111) lattices experienced an in-plane compressive strain along c-axis (due to increasing *d*-spacing), while β-Ta(002) lattices were subjected to an in-plane tensile strain. It is evident, that there is a tendency to unify *d*-spacing of both Ta and Cu, when $\Lambda$ decreases. On the other hand, for $\Lambda \geq 12$ nm, the *d*-spacing of Ta(002) increases while Cu(111) is almost constant. It means that the $\sigma$ in the Cu layers remains constant (almost zero), comparable to the $\sigma$ in the Cu monolayer film (according to *d*-spacing), while the $\sigma$ in the Ta layers shift from small compressive to high compressive with increasing layer thickness – see shift of the Ta (002) and Ta(004) diffraction peaks to lower $2\Theta$ angles in Fig. 1. The compressive $\sigma$ in Ta layers is gradually approaching to value of the Ta monolayer film, with increasing $\lambda_{Ta}$. A similar tendency was observed by J. J. Colin et al. [35] in Ta film, where increasing Ta thickness from 1 nm to 10 nm led to a shift in film $\sigma$ from tensile to compressive. The average $\sigma$ in the Ta/Cu NMM films was compressive and increased from −0.11 GPa to −0.67 GPa with increasing $\Lambda$. Moreover, the Ta/Cu films deposited at high $a_D$ exhibit higher $\sigma$. Due to the different crystal structures of Ta and Cu (tetragonal vs. face-centred cubic) and their high lattice mismatch ($a = 1.02$ nm and $b = 0.53$ nm) and ($a = 0.36$ nm), respectively, the resulting interfaces are expected to be strongly incoherent.

To gain detailed information about the structure and Ta/Cu interface of the films, with respect to similar XRD patterns of both series deposited

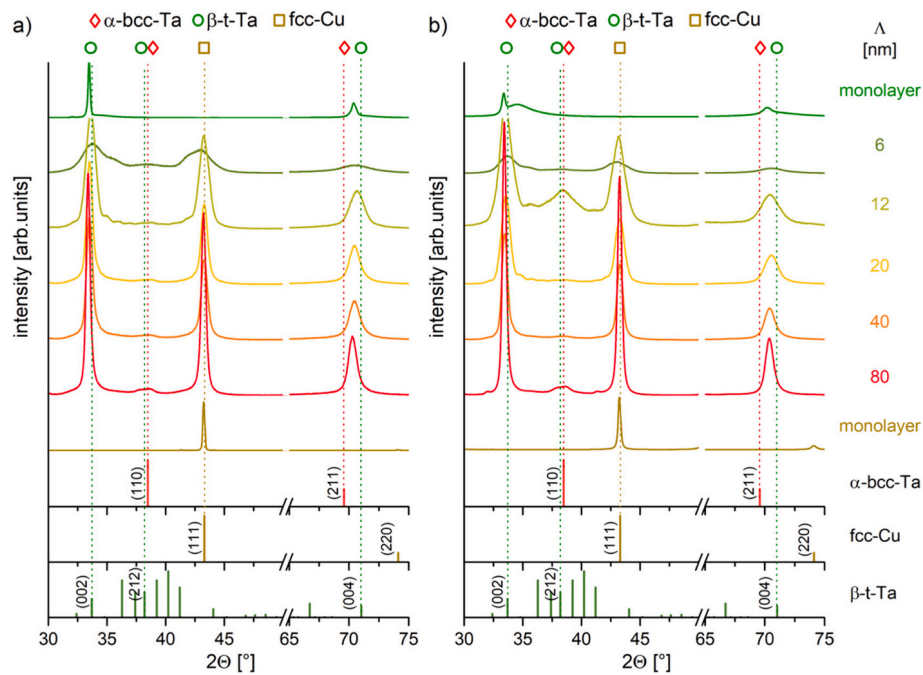

**Fig. 1.** XRD patterns of the as-deposited Ta/Cu NMMs with different periodicities ($\Lambda$), deposited at a low $a_D$ (a) and a high $a_D$ (b).

**Table 1**

Structural properties of the as-deposited Ta/Cu NMMs as a function of the periodicity $\Lambda = \lambda_{Ta} + \lambda_{Cu}$. Where $\lambda_{Ta}$ and $\lambda_{Cu}$ are the layer thicknesses, $a_D$ is the deposition rate, $D$ is the average crystallite size, and $\sigma$ is the average biaxial residual macro-stress in the films.

| Film | $a_{D\ Ta}$/ $a_{D\ Cu}$ [nm/min] | $\Lambda$ [nm] | $\lambda \approx \lambda_{Ta} \approx \lambda_{Cu}$ [nm] | $D_{Ta(002)}$ [nm] | $D_{Cu(111)}$ [nm] | d-spacing [nm] | | $\sigma$ [GPa] |
|---|---|---|---|---|---|---|---|---|
| | | | | | | Ta(002) | Cu(111) | |
| Cu | 11 | Monolayer | | – | ~41 | – | 0.2090 | ±0.02 |
| Ta | 10 | Monolayer | | ~35 | – | 0.2675 | – | −1.69 |
| Ta/Cu$_6$ | 10/11 | 6 | ~3 | ~4 | ~4 | 0.2645 | 0.2110 | −0.11 |
| Ta/Cu$_{12}$ | 10/11 | 12 | ~6 | ~6 | ~6 | 0.2650 | 0.2092 | −0.19 |
| Ta/Cu$_{20}$ | 10/11 | 20 | ~10 | ~8 | ~8 | 0.2663 | 0.2090 | −0.15 |
| Ta/Cu$_{40}$ | 10/11 | 40 | ~20 | ~12 | ~12 | 0.2671 | 0.2088 | −0.34 |
| Ta/Cu$_{80}$ | 10/11 | 80 | ~40 | ~18 | ~17 | 0.2681 | 0.2091 | −0.67 |
| Cu | 25 | Monolayer | | – | ~30 | – | 0.2091 | ±0.02 |
| Ta | 20 | Monolayer | | ~8 | – | 0.2970 | – | −2.08 |
| Ta/Cu$_6$ | 25/20 | 6 | ~3 | ~6 | ~4 | 0.2654 | 0.2104 | −0.29 |
| Ta/Cu$_{12}$ | 25/20 | 12 | ~6 | ~6 | ~7 | 0.2667 | 0.2095 | −0.32 |
| Ta/Cu$_{20}$ | 25/20 | 20 | ~10 | ~9 | ~9 | 0.2669 | 0.2089 | −0.27 |
| Ta/Cu$_{40}$ | 25/20 | 40 | ~20 | ~13 | ~13 | 0.2675 | 0.2089 | −0.55 |
| Ta/Cu$_{80}$ | 25/20 | 80 | ~40 | ~20 | ~18 | 0.2678 | 0.2090 | −0.62 |

at low and high $a_D$, two selected samples Ta/Cu$_6$ and Ta/Cu$_{20}$ deposited at a low $a_D$ were studied by means of TEM. Fig. 2 shows HAADF images and EDX mapping of selected NMMs samples. From HAADF image, the thickness of both Ta and Cu layers was about 3–4 nm for Ta/Cu$_6$, and 10 nm for Ta/Cu$_{20}$ sample, which matches the one expected in Table 1. In the case of the sample with the smallest period, a "woven" structure of the film was found. Such phenomena probably can be related to the fact that Cu growth follows the three-dimensional (3D) island (Volmer-Weber) [36–38] mode, and before the island coalescence occurs [39], the growth is stopped by deposition of next Ta layer on the top of Cu. The image of the Ta/Cu$_{20}$ sample (shown in Fig. 2a) indirectly confirms this assumption; insignificant inhomogeneity in thickness was found at the boundaries between Cu grains, where coalescence has not been completed fully [40]. One possible explanation of 3D growth of the Cu layer on Ta sublayer is its surface energy minimization. Cu and Ta are practically immiscible with positive mixing enthalpy of +12 kJ/mol [41] in the solid state, and therefore there is a lack of driving force for the formation of Ta—Cu compound (e.g., intermetallic). In other words, positive mixing enthalpy acts by repulsive forces between Ta and Cu atoms [31]. Particularly, the bond strength between Ta and Cu is very low. Thereafter, the strength between metallic bonds of Cu—Cu of the growing Cu layer starts to have a significant role that exceeds the bonding strength between Cu and Ta sublayer. The surface energy minimization during 3D Cu growth is responsible for the development of the (111) texture in the *fcc*-Cu (see Fig. 1) [42]. The occurrence of 3D Cu islands before their coalescence cause maximum tensile stress in the layer, which contribute to the lowering of the residual compressive macro-stress in the whole film [43]. The microstructure of the Ta/Cu$_6$ film exhibits a fiber-like nanocomposite, wherein the soft and ductile nanocrystalline *fcc*-Cu phase is interwoven with the hard and brittle nanocrystalline *t*-Ta phase.

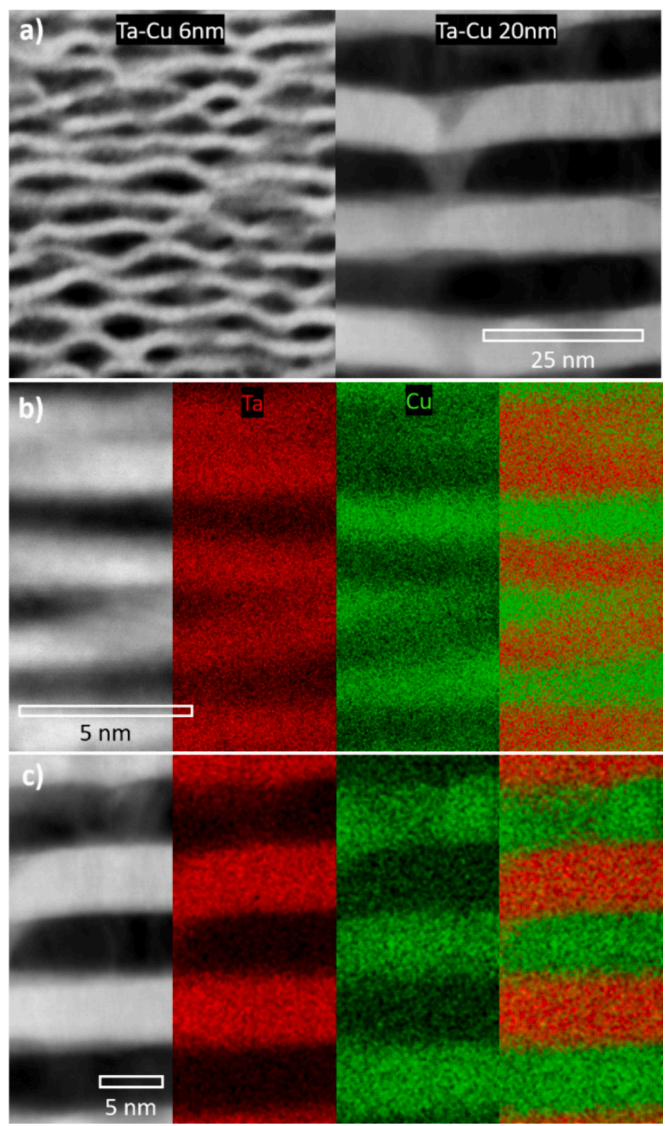

**Fig. 2.** (a) HAADF images of the Ta/Cu$_6$ and Ta/Cu$_{20}$ films, and EDX mapping of (b) the Ta/Cu$_6$ film, and of (c) the Ta/Cu$_{20}$ film.

EDX mapping (Fig. 2b, c) shows that the Ta/Cu interface of as-deposited samples is sharp, with discrete Cu and Ta elemental distributions and without any extensive interdiffusion [44]. Although magnetron sputtering is a non-equilibrium process, especially at a low $p_{Ar}$, and can cause Ta—Cu intermixing in several atomic layers during the deposition of the Ta with high energy, immiscibility clearly prevailed in this case.

Fig. 3a, b represents SAED patterns taken from the middle part of the films (diameter of the aperture ~250 μm). The blurry ring SAED pattern from the Ta/Cu$_6$ sample corresponds to the small size of the crystallites, whereas the pattern has intensity "hot spots" which means a preferential orientation of crystallites in the film. Following SAED pattern from the Ta/Cu$_{20}$ sample, a preferential orientation becomes more pronounced with an increase of a period $\Lambda$ (Fig. 3b). Integrated intensity profiles extracted from the SAED patterns are displayed in Fig. 3c. Identified peaks correspond to the planes (111), (002), (022), (113), (222), and

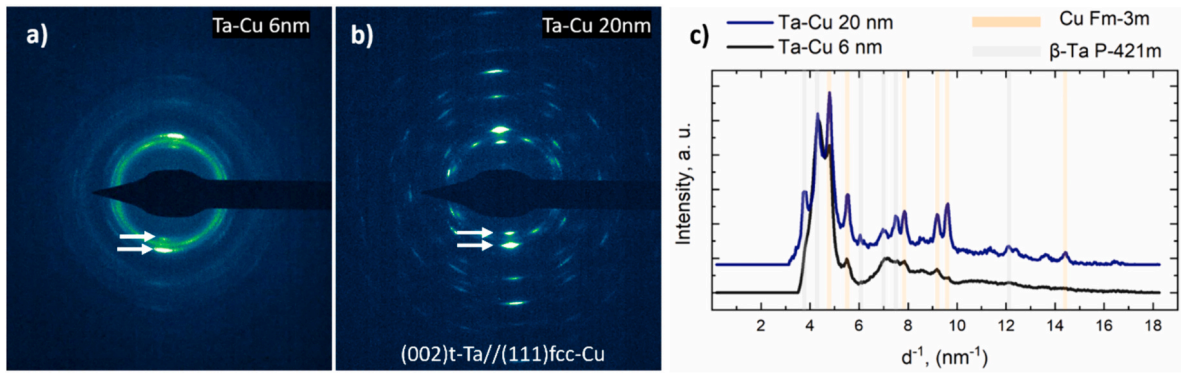

**Fig. 3.** SAED patterns taken from (a) the Ta/Cu₆ and (b) Ta/Cu₂₀ films deposited at a low $a_D$, and (c) the radial intensity profiles from these SAED patterns.

(115) of the *fcc*—Cu, as well as to the (002), (022), (123), (333), (004), (336) planes of β-Ta phase (on Fig. 3 c all the planes are marked from left to right). It is crucial to note that the (002) spot from β-Ta aligns with the (111) spot of *fcc*-Cu (marked by white arrows on Fig. 3 a,b), indicating that these planes are parallel to each other. The results of SAED are in good agreement with XRD data shown in Fig. 1a.

Fig. 4 shows HRTEM images of the Ta/Cu₆ and Ta/Cu₂₀ films deposited at a low $a_D$. In the case of Ta/Cu₆, it is observed that the Cu and Ta layers predominantly exhibit a monocrystalline structure, although instances of polycrystalline variants are also identified. The size of crystallites in a growth direction is limited by layer thickness, and in the direction parallel to the substrate surface they can reach a few tens of nanometers. Only a few regions in the Ta layers are highly disordered and exhibit amorphous-like contrast, which may also contribute to the halo at $2\Theta = 38°$ in the XRD patterns (see Fig. 1). The HRTEM cross-section image shows that at a low periodicity of 6 nm, the Cu layers are not continuous and have a fiber-like structure (due to 3D island growth) where the Cu crystallites are encapsulated in the β-Ta phase. As the period increases (Ta/Cu₂₀), the Cu forms a continuous layer, and each layer reveals polycrystalline structure. Scanning of both Ta/Cu₆ and Ta/Cu₂₀ samples using the NBD technique shows that the absolute majority of Cu is cubic structure along with [110] zone axes parallel to an incident beam (Fig. 4b). The NBD patterns from Ta layers are characterized as a tetragonal structure along with [110], [120], [100] zone axis, between these orientations no preferential one was found (Fig. 4c, e). HRTEM imaging reveals that a boundary between t-Ta layers and *fcc*-Cu layers also has diffuse and disordered regions. The thickness of the transition region between *fcc*-Cu and t-Ta grains is found to be inhomogeneous, ranging from 1 atomic layer to 1.5 nm. We assume these amorphous regions are composed of both Cu and Ta atoms as reported elsewhere [13,16,31]. The sharp interfaces with atomically-ordered structures and/or uncommon interfacial transition zone [45–47], that are typical of immiscible alloys, were not detected in our case. Important to highlight, the growth direction of both Cu and Ta layers was found to be along <001>β-Ta || <111>Cu directions in all cases, and schematic plot of the "perfect" atomic arrangement of the crystal junction line is presented in Fig. 4f.

Fig. 6 shows cross-sectional SEM images of all Ta/Cu NMM films with a different $\Lambda$. For Ta/Cu with $\Lambda \leq 12$ nm, a featureless and columnar-less structure without recognizable layers is observed, as shown in Fig. 6a,b, while a layered structure is observed for Ta/Cu with $\Lambda \geq 20$ nm, which is more pronounced for the 80 nm, as shown in Fig. 6c-d. It can be observed that incoherent and/or amorphous Ta—Cu interface and its increasing fraction (with decreasing periodicity) prevent the continuous growth of the columnar microstructure throughout the film thickness. The bright interface between the Si(100) substrate and Ta/Cu NMM is due to the non-conducting amorphous Si₃N₄ seed layer with a thickness of 50 nm, which suppresses epitaxial growth from the substrate.

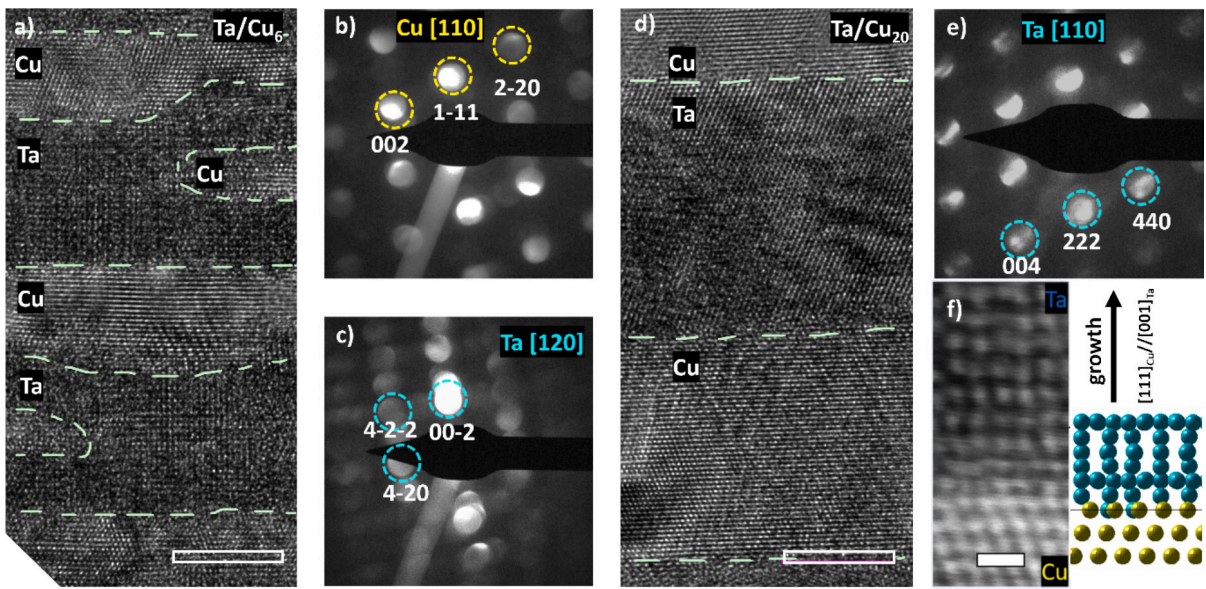

**Fig. 4.** HRTEM images of (a) Ta/Cu₆ and (d) Ta/Cu₂₀ films (with 5 nm scale bar); NBD patterns taken from (b) Cu and (c, e) Ta layers, and (f) the BF STEM image of the interface between two layers (5 Å scale bar) with modeled "perfect" interface.

### 3.2. Mechanical properties

#### 3.2.1. Hardness, elastic modulus, and stress

The mechanical properties ($H$, $\sigma$, $E^*$, and elastic recovery $W_e$) of the as-deposited Ta/Cu NMMs as a function of $a_D$ (low and high) and $\Lambda$ are shown in Fig. 7a,b and summarized in Table 2. From Fig. 6a, it can be seen that the evolution of $H$ for Ta/Cu films deposited at low $a_D$ is monotonous, gradually increasing from 6.5 GPa at $\Lambda = 80$ nm to reach a plateau at $H = 7.5$ GPa ($\sigma_f = 2.8$ GPa) at $\Lambda \leq 12$ nm, while at high $a_D$ the evolution of $H$ is strongly non-monotonic, decreasing from 6.6 GPa to 4.9 GPa as $\Lambda$ decreases from 80 nm to 20 nm and then increasing rapidly to a plateau at $H_{max} = 8.2$ GPa ($\sigma_{f\,max} = 3.0$ GPa) at $\Lambda \leq 12$ nm. The increase in $H$ for Ta/Cu NMMs with $\Lambda \leq 12$ nm can be attributed to low $D$ (4 and 7 nm), where approximately 3 and 1 dislocations may be formed within Ta and Cu crystallites, respectively [48]. It was found that the higher the $a_D$, the higher the $H$. The mechanisms of strengthening are investigated and discussed in the last section.

To assess the hardness of the NMMs with superlattice, here of the Ta/Cu films, the rule-of-mixture (ROM) is used:

$$H_{ROM} = V_{Ta}\,H_{Ta} + V_{Cu}H_{Cu} \qquad (1)$$

Where $V_{Ta}$ and $V_{Cu}$ and $H_{Ta}$ and $H_{Cu}$ are the volume fractions and hardness of Ta and Cu layers, respectively. The $H$ and $E^*$ values of the 1.5 μm thick tantalum and copper monolayers used for the $H_{ROM}$ calculation are given in Table 2. Since the values of $V_{Ta}$ and $V_{Cu}$ are almost the same, $V_{Cu} \approx V_{Ta} = 0.5$ ($\lambda_{Ta} \approx \lambda_{Cu}$ see Table 1), the $H_{ROM}$ is equal to 9.5 GPa and the flow strength calculated form $H_{ROM}$ is $\sigma_{f\,ROM} = 3.52$ GPa. A comparison of the experimentally measured $H$ with the calculated $H_{ROM}$ is shown in Fig. 6a. We can see that $H_{ROM}$ is notably greater than the observed $H$ of all as-deposited Ta/Cu NMMs. This can be explained by the fact that the $H$ of highly compressively stressed ($-1.69$ GPa and $- 2.08$ GPa, see Table 1) Ta films deposited at low and high $a_D$, respectively, were used to calculate the $H_{ROM}$, which overestimate the real values. The $H_{max}$ of Ta/Cu$_6$ was found to correlate well with the $H$ of a Ta—Cu alloy with a similar volume fraction (with $\sim$50 at.% Cu) prepared by W. Quin et al. [49]. They also observed that the hardness of Ta—Cu alloy with $\sim$50 at. % Cu is lower than that of $H_{ROM}$ – the same as that of our deposited Ta/Cu NMMs.

All the as-deposited Ta/Cu NMMs have residual compressive macro-stress ranging from $-0.11$ GPa to $-0.67$ GPa (see Fig. 6a and Table 1), which usually results in a hardness enhancement. The compressive $\sigma$ significantly affects the $H$ of Ta and Ta/Cu NMM films. The $H$ of Ta/Cu$_6$ and Ta/Cu$_{12}$ is higher for the films deposited at high $a_D$. One might expect the decrease in $H$ with increasing $\Lambda$ to be even higher when the $\sigma$ approaches to zero for Ta/Cu$_{40}$ and Ta/Cu$_{80}$. Recall that the origin of the compressive $\sigma$ is mainly attributed to bombardment of the growing films by high-energy fast neutrals (mainly Ta) and backscattered Ar neutrals [28,50]. The resulting $\sigma$ in both Ta/Cu$_{80}$ nearly matches the calculated average $\sigma$ of $-0.85$ GPa from as-deposited Ta and Cu monolayers with $\sigma$ of $-1.7$ GPa and $\pm$ 0.02 GPa, respectively. The decrease in the compressive $\sigma$ with the $\Lambda$ in the Ta/Cu NMMs can be explained by the decrease in $\lambda_{Ta}$ [51,52].

#### 3.2.2. Deformation behavior – creep and high-load tests

The elasto-plastic mechanical properties of the as-deposited Ta/Cu NMMs are represented by the elastic recovery $W_e$ (shown in Fig. 6b) evaluated from the nanoindentation loading-unloading curves. The $W_e$ was calculated as a ratio of elastic to total indentation depth. If $W_e > 50\%$, the material is deformed rather elastically, while if $W_e < 50\%$, the material is deformed rather plastically. In our system Ta/Cu, the hard ($H = 17$–18 GPa) β-Ta monolayer film exhibits a remarkably high elastic recovery of 67–75%, while the soft ($H = 1.6$–3.0 GPa) fcc-Cu monolayer film exhibits a low elastic recovery of 9–12%. This means that the Cu layer should be mainly responsible for the plastic deformation of Ta/Cu NMMs. Fig. 7b shows that the $W_e$ of Ta/Cu NMMs corresponds to the average values of the Ta and Cu monolayers, see Table 2. The $W_e$ increases with an increasing number of the Ta—Cu interfaces, i.e. with decreasing $\Lambda$. The increase in $W_e$ is more pronounced when a Cu "woven" nanolaminate microstructure forms in Ta/Cu (see Figs. 2a and 3a) with $\Lambda \leq 12$ nm. Thus, it can be concluded that increasing the interface density increases the resistance to plastic deformation of the Ta/Cu NMMs. To assess and understand the deformation behavior of the Ta/Cu NMMs, indentation tests under high load were performed.

The surface SEM images of the imprints in the as-deposited Ta/Cu NMMs, and the cross-sectional HAADF image of imprint in Ta/Cu$_6$ and Ta/Cu$_{20}$ NMMs are shown in Figs. 7 and 8, respectively. The indentation was carried out by the Berkovich diamond indenter at a high load of 200 mN, with a maximum indentation depth of 1.15–1.17 μm, i.e., between 58% and 62% of the film thickness. The imprints after indentation were cross sectioned by FIB technique for further S/TEM analysis (Fig. A1). In contrast to the indentation at a low load of 4 mN, the formation of circular cracks under high load was observed around the imprints. No radial cracks were observed in the corners of the imprints. The absence of the radial cracks can be explained by the following facts: (1) the high interface density suppresses the columnar growth in the NMM films (see the cross-sectional SEM morphology in Fig. 5, which acts as the weakest place for crack propagation [53]; (2) the compressive macro-stress in the as-deposited NMM films (see Table 1 and Fig. 7a which acts against the possible crack propagation by their closing [54]; (3) multilayer toughening [54] where Ta and Cu exhibit different $H$, $E^*$, $W_e$ and compressive stress, alternating elastically deformed β-Ta layers with high toughness [8], and plastically deformed Cu layers with high ductility, where both prevent the formation of crack.

From the cross-section STEM images of the imprints shown in Fig. 8, it can be seen that the circular cracks represents pile-ups (indicated by arrows in Figs. 8b and A2) caused by shear band (SB) formation. The SBs are also observed as kinks (pop-ins) in the loading curves (see Fig. A3 in Appendix). SBs formation was observed in all Ta/Cu NMMs, tested at a high load of 200 mN, while SBs were not observed at a low load of 4 mN, where the indenter depth was low between 0.13 and 0.15 μm (see Table 2). For Ta/Cu with $\Lambda = 6$ nm, there are no recognizable kinks in the loading curve, which correlates with the indent morphology (Fig. 7a), where there are only small circular cracks compared to the others. However, on the HAADF cross-section image some minor traces of SB formation are visible in the case of Ta/Cu$_6$ sample, but these traces are well less pronounced in comparison with Ta/Cu$_{20}$ sample. This difference can be explained by the higher interface density compared to other samples with $\Lambda > 6$ nm. the woven nanostructure where Cu is encapsulated in Ta, high elastic recovery of Ta up to 75%, indicating an elasto-plastic deformation behavior. In general, for this system, a formation of SB can be described as follows: during indentation of the Ta/Cu NMM film, the stress energy caused by dislocations (expected dislocation propagation mainly in Cu) is stored near the grain boundaries (GB) before their release.

**Table 2**
Mechanical properties of the as-deposited Ta/Cu NMMs. Where $d_i$ is the average maximum indentation depth, $W_e = d_{el}/d_i$ is the elastic recovery, $d_{el}$ is the elastic depth, and $\sigma_f = H/\alpha$ is the flow strength.

|  | Film | $H$ | $E^*$ | $d_i$ | $W_e$ | $\sigma_f$ |
|---|---|---|---|---|---|---|
|  |  | [GPa] | [GPa] | [nm] | [%] | [GPa] |
| Low $a_D$ | Cu | 1.6 ± 0.1 | 129 | 118 | 9 | 0.6 |
|  | Ta | 17.6 ± 0.1 | 191 | 114 | 67 | 6.5 |
|  | Ta/Cu$_6$ | 7.5 ± 0.2 | 142 | 145 | 44 | 2.8 |
|  | Ta/Cu$_{12}$ | 7.5 ± 0.1 | 155 | 146 | 41 | 2.8 |
|  | Ta/Cu$_{20}$ | 7.0 ± 0.1 | 156 | 150 | 38 | 2.6 |
|  | Ta/Cu$_{40}$ | 6.9 ± 0.1 | 161 | 151 | 37 | 2.5 |
|  | Ta/Cu$_{80}$ | 6.5 ± 0.1 | 161 | 133 | 38 | 2.4 |
| High $a_D$ | Cu | 3.05 ± 0.04 | 174 | 109 | 12 | 1.1 |
|  | Ta | 17.0 ± 0.2 | 193 | 72 | 75 | 6.3 |
|  | Ta/Cu$_6$ | 8.2 ± 0.1 | 154 | 104 | 54 | 3.0 |
|  | Ta/Cu$_{12}$ | 8.1 ± 0.1 | 155 | 105 | 53 | 3.0 |
|  | Ta/Cu$_{20}$ | 4.9 ± 0.1 | 140 | 138 | 34 | 1.8 |
|  | Ta/Cu$_{40}$ | 6.4 ± 0.1 | 161 | 118 | 39 | 2.4 |
|  | Ta/Cu$_{80}$ | 6.6 ± 0.2 | 162 | 116 | 44 | 2.5 |

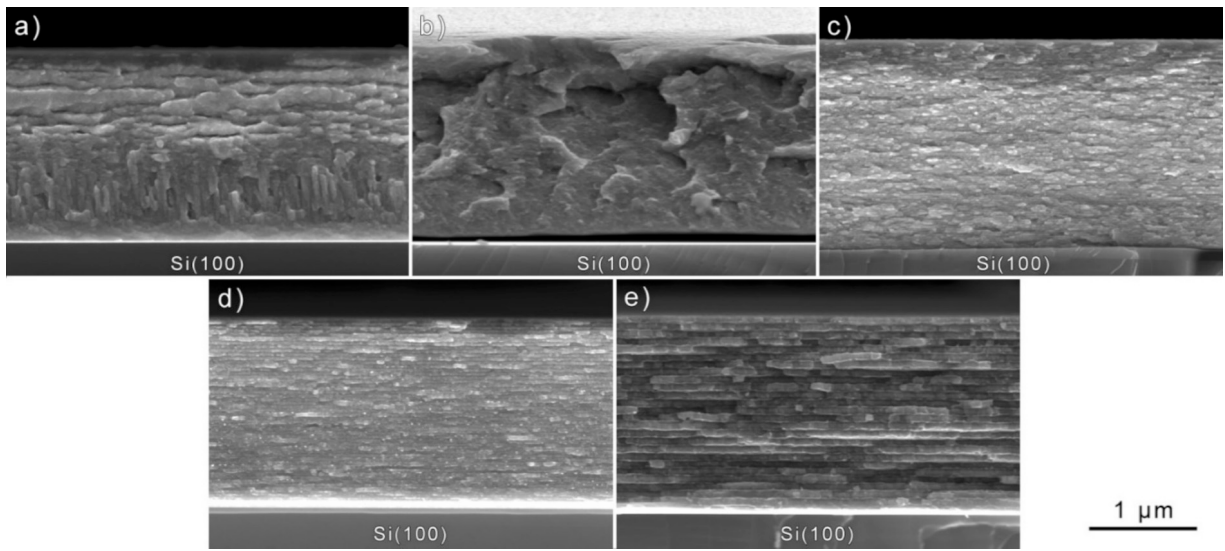

**Fig. 5.** Cross-sectional SEM images of a morphology of the as-deposited Ta/Cu NMMs at a low deposition rate as a function of the periodicity $\Lambda$: (a) 6 nm, (b) 12 nm, (c) 20 nm, (d) 40 nm, and (e) 80 nm.

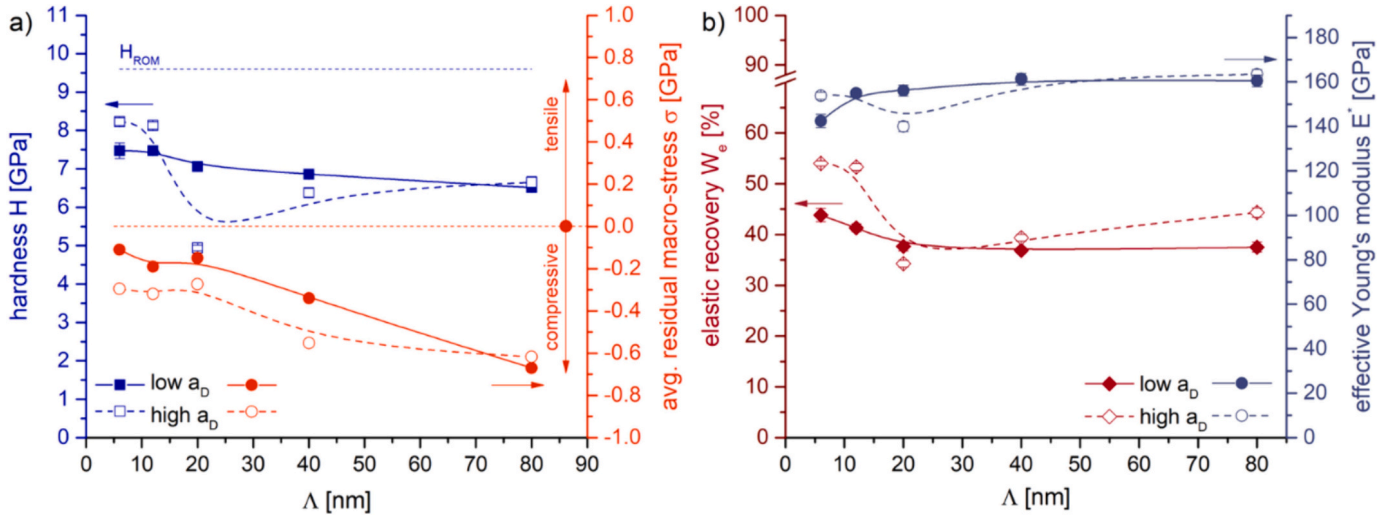

**Fig. 6.** Mechanical properties (a) $H$ and $\sigma$, and (b) $W_e$ and $E^*$ of the as-deposited Ta/Cu NMMs as a function of the $\Lambda$ deposited at low and high deposition rate $a_D$.

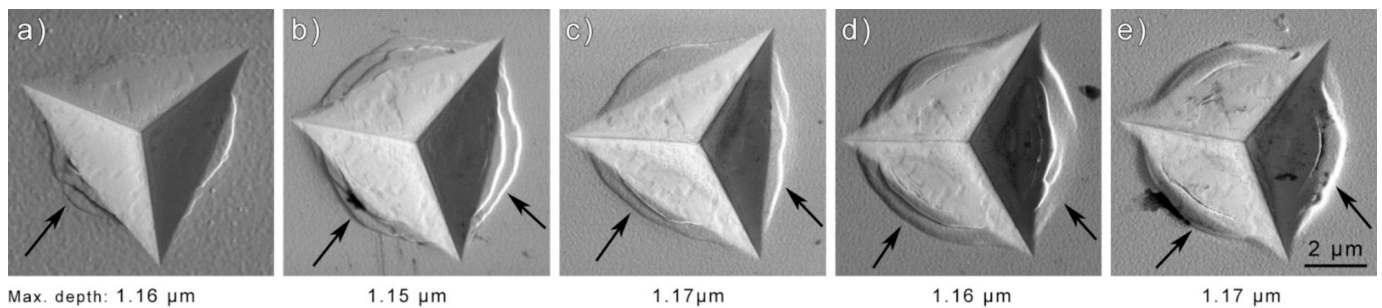

**Fig. 7.** SEM images depicting indentation imprints on the as-deposited Ta/Cu NMMs deposited at a low $a_D$ as a function of periodicity $\Lambda$: (a) 6 nm, (b) 12 nm, (c) 20 nm, (d) 40 nm, and (e) 80 nm The arrows point at the surface shear band (SB) features.

Here, the GB and/or the Ta/Cu interface can act as a barrier against dislocation motion and propagation. The amount of stored energy causes local strengthening (when dislocations pile-up), which represents local interfacial energy. After overcoming this energy barrier, dislocations are transmitted across the GB within the layer and/or Ta/Cu interface, the stress energy is released, and local softening occurs. This leads to irreversible plastic deformation; shear across the interfaces – pop-in formation. F. Wang et al. [11] observed SBs formation in their Ta/Cu films except for $\lambda > 30$ nm – this contrasts with our observations. Further studies on CuTa/Cu or Ta/Cu multilayers with a total Cu content ranging from 47 to 73 at. % shows the formation of SBs [49,55], while at a total Cu content lower than that 40 at.%, the formation of SBs was not observed [56]. Formation of the SBs is a typical deformation process of metallic glasses (bulk and/or thin film) with an amorphous structure, and represents their catastrophic failure [57,58].

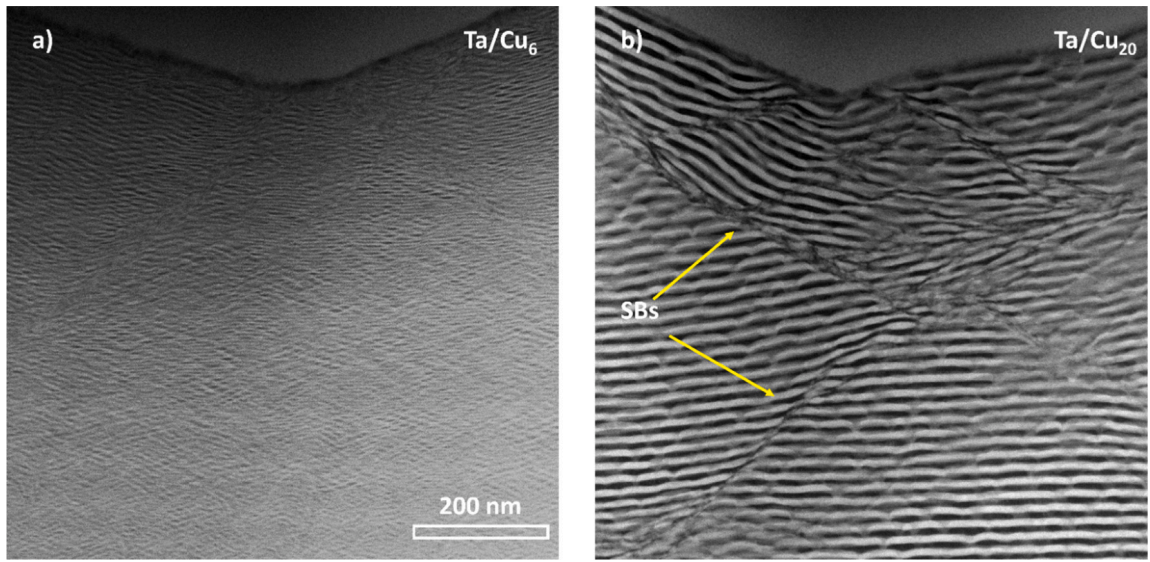

**Fig. 8.** Cross-section HAADF images of the deformed (a) Ta/Cu$_6$ and (b) Ta/Cu$_{20}$ NMMs deposited at a low $a_\mathrm{D}$. The scale bar remains consistent for both (a) and (b).

TEM studies shed some light on deformation behavior of NMMs. The undeformed Ta/Cu$_{20}$ films exhibit a low density of {111} growth nanotwins, as illustrated in Fig. 9a. These nanotwins are heterogeneously distributed within the grains, some containing several twin boundaries. The FFT pattern obtained from a boxed region oriented along a [110] zone axis reveals the characteristic superposition of two mirrored [110] patterns concerning the {111} plane. Most growth twins are parallel to the film surface, confirming the presence of crystallographic texture in the growth direction. The twin boundaries (TBs) are predominantly incoherent, indicating the accumulation of Shockley partial dislocations (SPDs) associated with stacking faults (Fig. 9a inset). Additionally, the Ta and Cu nanolayers appear devoid of dislocation cell structures or networks. Upon analyzing TEM images from a lamella prepared from an imprint cross-section after deformation, microstructural changes are evident. In the TEM images (Fig. 9b and A5), the appearance of contrast suggests the presence of screw dislocations in the Cu layers. HRTEM images reveal a dramatic decrease in TB coherency after indentation compared to the as-deposited film. This reduction in coherency might be attributed to the interaction of dislocations with existing growth nanotwins. The TB exhibits steps related to the presence of several SPDs in the twinning plane. SPDs can glide on the {111} plane, accompanied by the creation of stacking faults. Frank dislocations (FD) are also identified at the TBs. Since a Frank partial is an edge dislocation perpendicular to {111}, its source is not confined to a single gliding plane but spans from gliding to above or below the atomic plane. The inset in Fig. 9b exhibits a HRTEM image capturing the distorted region of the *fcc*—Cu. It specifically illustrates the out-plane displacement of atoms within the {111} plane, giving a visualization of the Frank partial formation. The observations conclusively affirm that plastic deformation is facilitated through the nucleation and glide of dislocations. The notable presence of various impediments to dislocation motion, such as grain boundaries, TBs, SFs, and the inclusion of Ta layers, significantly contributes to the observed high strength characteristics of the NMM material.

To assess the ductility and deformation mechanism in the as-deposited Ta/Cu NMM films, the strain rate sensitivity $m$ and activation volume $V^*$ were evaluated from the indentation creep test under a constant load for a constant time, respectively. Creep displacement vs. time data (see Fig. A4 in the supplementary data) were evaluated by the procedure described in [15,22]. Both $m$ and $V^*$ are used to describe thermodynamics and kinetics of material plastic deformation [59]. The values $m$ and $V^*$ as a function of $\Lambda$ are plotted in Fig. 10. The figure shows that an increasing (decreasing) tendency for $m$ ($V^*$) with $\Lambda$ to $\Lambda = 20$ nm, and with further increasing $\Lambda$ the tendency for $m$ ($V^*$) is vice-versa. In the case of low $a_\mathrm{D}$, the highest (lowest) $m = 0.024$ ($V^* = 11.2$ b$^3$), and for high $a_\mathrm{D}$ $m = 0.025$ ($V^* = 13.4$ b$^3$) values are at $\Lambda = 20$ nm. It is well known that $m$ is highly sensitive to the average grain size. Usually, for *fcc* metals, the $m$ increases with decreasing the grain (crystallite) size, while for bcc metals, it is vice-versa – the $m$ decreases with decreasing the grain (crystallite) size [60]. In the case of hcp metals, this direct trend is not unambiguous [60]. Q. Zhou et al. [60] found out that the layer thickness of both constituents (*bcc*-Ta and *fcc*—Cu) ranging from 2 to 200 nm has no significant effect on $m$, which was constant $m = 0.052$, and more than two times higher than our estimated average value $m_\mathrm{avg} = 0.019$ through $\Lambda$. Further, they found out that with increasing *fcc*-Cu/bcc-Ta thickness ratio from 0.17 to 0.83, the $m$ increases from 0.035 to 0.055. It can be accompanied by an increase in ductility because the Cu phase is more ductile than that of the Ta phase, even more than the ductile bcc-Ta phase. Therefore, based on that calculation the Ta/Cu$_{20}$ NMM with the highest $m$ should exhibit enhanced ductility, see Fig. 10. This result is also supported by the fact, that Ta/Cu$_{20}$ films (at a low and high $a_\mathrm{D}$) exhibit lowest elastic recovery (38% and 34%), i.e., highest plastic deformation. A quantitively different value of $m = 0.005$ was estimated in the Ta—Cu alloy with 50 at.% by E. Reaker et al. [61]. Nano-twinned and nanocrystalline *fcc*-Cu shows increased $m$ and reduced $V^*$, which results in increased strength without severely compromising ductility [62].

In the case of $V^*$, it was suggested that the $V^* < 1$ b$^3$ is typical for GB sliding or GB diffusion mediated creep (Coble creep), the $V^*$ ranging of 1–100 b$^3$ typically involves cross-slip of screw dislocations (Fig. A5), and/or at $V^* \sim 1$–10 b$^3$ dislocation emission from interfaces and GBs occurs [63]. A typical bulk dislocation source, such as the Frank-Read source, is between 100 and 1000 b$^3$, leading to the commonly observed rate-insensitive plastic yield in large crystals at room temperature [64]. Under these assumptions where the $V^*$ is ranging from 11 to 20 b$^3$, is assumed a mixed deformation mechanism: dislocations emission from the incoherent Ta/Cu interfaces and/or GBs of each constituent and the formation of screw dislocations inside the grains. Obstruction of the dislocation formation (pileups) in small crystallites of 4–5 nm (see Table 1) could be the physical reason behind the enhanced hardness (strength) of 7.5 GPa (2.8 GPa) in the Ta/Cu$_6$ NMM deposited at low $a_D$. But on the other hand, a decrease in the interface density and/or increase in soft Cu layer thickness leads to a reduction of dislocation obstacles and then to softening of the Ta/Cu NMMs compared to their ROM values. The deformation mechanisms that describe deformation within Ta and Cu layer and across the Ta—Cu interface by using refined confined layers slip model and interface barrier slip model, respectively, is discussed in following section.

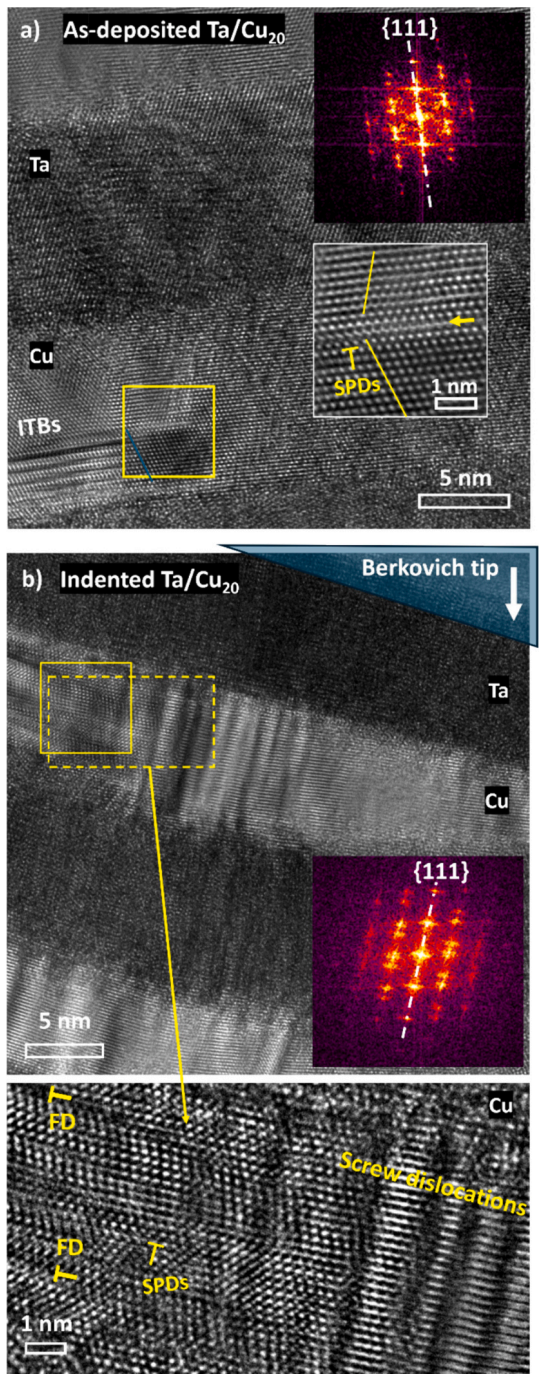

**Fig. 9.** Cross-section HRTEM images of the deformed (a) Ta/Cu$_6$ and (b) Ta/Cu$_{20}$ NMMs deposited at a low $a_D$.

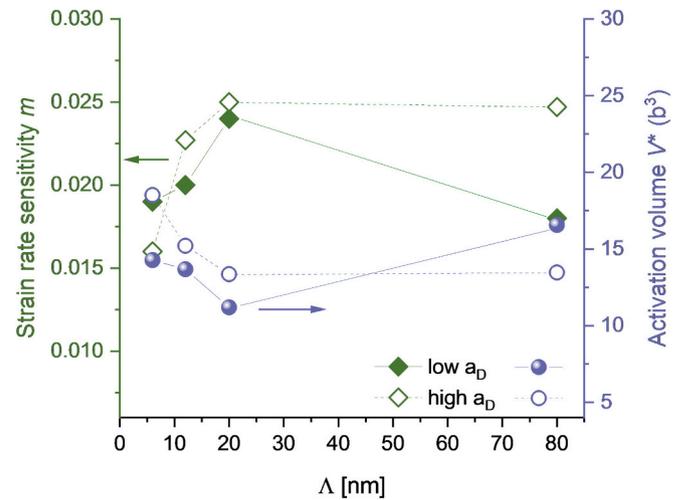

**Fig. 10.** Strain rate sensitivity and activation volume of the Ta/Cu NMMs as a function of the periodicities $\Lambda$. The evolution of the $m$ and $V^*$ is shown in the supplementary data.

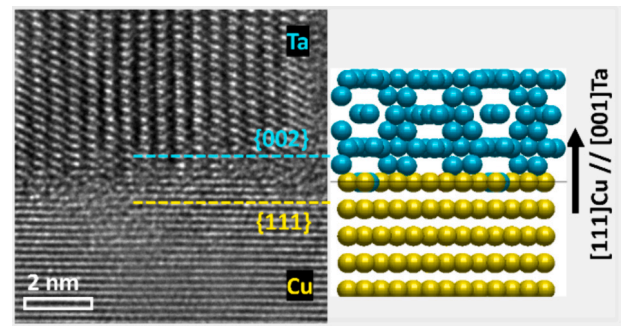

**Fig. 11.** HRTEM image on the interface in Ta/Cu$_{20}$ sample and model of a "perfect" interface.

### 3.2.3. Strengthening mechanisms

When individual layer thickness $\lambda$ (in multilayer system $\lambda_{Me1}/\lambda_{Me2}$) is greater than ~50 nm (100 nm), the strengthening effect originating from dislocation pile-up at GBs typically follows Hall-Petch relation, that the hardness (flow strength) changes with $\lambda^{-1/2}$ [65–67].

When the layer thickness falls within the range of 5 nm < $\lambda$ < 50 nm (with a maximum of 100 nm), the refined confined layer slip (rCLS) model becomes applicable [9]. This model illustrates the confined motion of a dislocation between two interfaces, separated by nanometers. Extensive investigations have been conducted over the years on *fcc/fcc* and *fcc/bcc* NMMs. In some instances, the interfaces were characterized using Kurdjumov–Sachs and/or Nishiyama-Wasserman orientation relationships to describe the configuration of slip systems. For instance, atomic-scale modeling of dislocation nucleation on interfaces can be employed to determine the orientation relationship. The results suggest that even minimal changes in the interface structure of NMMs can have a significant impact on the activated slip systems in adjacent crystals [68]. Furthermore, twinning structures at the interface of materials with low stacking fault energy significantly influence their strength. However, detailed information on *fcc/tetragonal* NMMs is currently lacking. This underscores a gap in our understanding of the mechanical behavior and characteristics of such materials, emphasizing the need for further research and exploration in this specific domain [69]. In this context, it's important to recognize that a *fcc/tetragonal* system may exhibit distinct behavior due to their different crystal structures and fewer slip systems. Thus, it is crucial to emphasize that the strengthening mechanism of incoherent interfaces in nanolaminates is exceedingly complex. Detailed investigations into the slip systems between adjacent *fcc/tetragonal* layers and the effects of the "woven" structure, as seen in the Ta/Cu$_6$ sample, are beyond the scope of the present work. It is imperative to highlight that the understanding of these intricate mechanisms requires further dedicated studies and verifications. The current work serves as an initial attempt to apply previously established equations to materials with a specific copper/tantalum interface structure. Readers should recognize the tentative nature of the following results, keeping in mind the need for future investigations to refine and deepen our comprehension of these complex strengthening mechanisms.

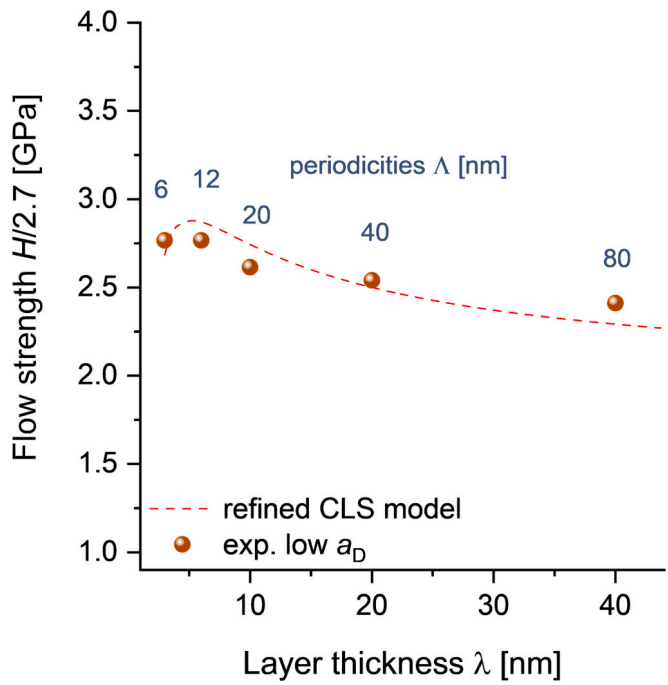

**Fig. 12.** Flow strength of the Ta/Cu NMMs deposited at a low $a_D$ as a function of the layer thickness $\lambda \sim \Lambda/2$. Calculation from the refined CLS model is depicted.

Our structural studies show that $\{002\}_{\beta\text{-Ta}} \parallel \{111\}_{Cu} \parallel$interface and $\langle 001 \rangle_{\beta\text{-Ta}} \parallel \langle 111 \rangle_{Cu}$ (Fig. 11), the interface is diffuse. The HRTEM image of the typical interface in the synthesized Ta/Cu films and possible "perfect" configuration of unit cells if given in Fig. 11.

The flow strength $\sigma_f$ of the as-deposited Ta/Cu NMM films deposited at a low $a_D$ as a function of layer thickness $\lambda$ ($\lambda_{Cu} \approx \lambda_{Ta}$) is shown in Fig. 12. The figure shows that with decreasing the $\lambda$, the $\sigma_f$ gradually increases and reaches a plateau value of 2.8 GPa at the $\lambda \le 6$ nm. Thanks to the small layer thickness and corresponding periodicities in the as-deposited Ta/Cu NMMs with $\lambda = 3$–40 nm, the $\sigma_f$, as well as the deformation mechanism, can be described by the rCLS model. Due to the soft Cu $H_{Cu} = 1.6$ GPa ($\sigma_f$ $_{Cu} = 0.6$–1.1 GPa) layer with the predominant plastic flow, see low $W_{e\ Cu} = 9\%$ compared to $W_{e\ Ta} = 67\%$, over the tough and hard Ta $H_{Ta} = 17.5$ GPa ($\sigma_f$ $_{Ta} = 6.3$–6.5 GPa) layer, we assume, that within the Cu layers the plastic deformation mostly occurs. Moreover, we must consider the significant contribution of the Ta layer hardness to the average measured hardness (strength). Thus, the stress (strength) required for single dislocation propagation in a confined Cu layer can be calculated by:

$$\sigma_{cls} = M \frac{\mu^* b_{Cu}}{8\pi\lambda'} \frac{4-\nu}{1-\nu} ln\left(\frac{\alpha\lambda'}{b_{Cu}}\right) + \frac{\mu^* b_{Cu}}{h(1+\nu)} + \frac{\gamma_{SF\ Cu}}{b_{Cu}} - \frac{f}{\lambda} \tag{2}$$

where the first term represents the stress needed to bow out an Orowan-type dislocation to propagate, the second term represents the dislocation-dislocation interactions during CLS, the third term is stress due to stacking faults behind the gliding, and the last term represents the interface stress contribution. In Eq. (2), $M$ is the Taylor factor, $\mu^*$ is the effective shear modulus of the Ta/Cu NMM calculated by combining shear moduli and volume fractions of the constituent elements as $\mu_{Ta} \cdot \mu_{Cu}/(V_{Ta} \cdot \mu_{Ta} + V_{Cu} \cdot \mu_{Cu})$, $b$ is the in-plane Burgers vector, $\lambda'$ is the layer thickness parallel to the glide plane ($\lambda' = \lambda/\sin \varphi$; $\varphi$ is the angle between the slip plane and the interface), $\nu$ is the Poisson ration, $\alpha$ represents the core cut-off (fitting) parameter, $h = b_{Cu}/\varepsilon$ is the spacing of the interface dislocation array (fitting parameter), $\varepsilon$ is the in-plane plastic strain, $\gamma_{SF}$ is the stacking fault energy, and $f$ is the characteristic interface stress. Substitution of $M = 3.06$, $\mu_{Ta} = 69$ GPa, $\mu_{Cu} = 48$ GPa, $b_{Cu} = 0.15$ nm, sin $\varphi = \sin 60° = (\sqrt{3})/2$, $\nu_{Cu} = 0.34$, $\nu_{Ta} = 0.35$, $\gamma_{SF}$ $_{Cu} = 0.045$ J/m$^2$, $f = 2$ J/m$^2$, and fitted parameters $\alpha = 0.103$ and $h = 8$ nm into Eq. (2), the flow strength of the as-deposited Ta/Cu NMMs can be calculated as a function of layer thickness, see dashed red line in Fig. 12. The rCLS fits well experimentally measured strength ($H/2.7$), see Fig. 12. The rCLS model confirms that the deformation of the Ta/Cu NMMs occurs mainly within the soft Cu layers, and strength can be calculated from effective shear modulus and dislocation propagation within the Cu layers. If we go more into the details, we find out that for flow strength depicted in Fig. 12, the in-plane spacing between glide dislocation loops (modeled from rCLS) is around $h = 8$ nm which corresponds to plastic strain $\varepsilon = 2.7\%$.

Considering all experimental results and the fact that the Ta layer is significantly harder than the Cu layer, we can conclude that the soft and ductile Cu layer is mainly responsible for the deformation of the Ta/Cu NMMs system according to the rCLS model. We assume that when Ta/Cu films with a high Cu volume fraction of ~50% [17] ($\Lambda = 2 \times \lambda$; where $\lambda_{Ta} \approx \lambda_{Cu}$) are deformed (in our case by indentation), an in-plane compression occurs, Cu layers undergo plastic deformation, while Ta layers undergo rather elastic deformation. After the critical stress is applied, dislocations within the small crystallites (grains that have approx. the same size as $\lambda$ in the growth direction, while in the in-plane direction their size can be several times larger) of the Cu layer pile-up up to a critical value (when, with respect to their spacing, there is no room for the formation of new dislocations), the flow strength of Ta/Cu NMMs increases rapidly. Then the flow strength of the softer constituents start to play an important role. Once the energy barrier, i.e. the flow strength of Cu and Ta/Cu interlayer, is overcome, strong plastic deformation occurs, which manifests itself as a catastrophic failure of the entire Ta/Cu NMM film (i.e. shear bands are formed).

## 4. Conclusions

All the collected findings on Ta/Cu nanoscale metallic multilayers sputtered at low and high deposition rates using unbalanced magnetron system with periodicities $\Lambda = 6$, 12, 20, 40, and 80 nm with the same layer thicknesses ($\lambda_{Ta} \approx \lambda_{Cu}$) are summarized as follows:

- XRD and (S)TEM results show formation of NMM films consisted of tetragonal β-Ta and face-centred cubic Cu immiscible layers, where $\{002\}_{\beta\text{-Ta}} \parallel \{111\}_{Cu} \parallel$interface and $< 001 >_{\beta\text{-Ta}} \parallel < 111 >_{Cu}$.
- The highest hardness and elastic recovery were observed for Ta/Cu NMM film with the lowest periodicity ($\Lambda = 6$ nm) of 7.5 GPa (8.2 GPa), and 41% (54%) for low (high) deposition rate, respectively.
- Calculation of the activation volume reveals that plastic deformation in the Ta/Cu NMM films occurs through the emission of dislocations from the Ta/Cu interfaces, grain boundaries, and their formation within the grains. The strain rate sensitivity model suggest that Ta/Cu$_{20}$ should demonstrate the highest ductility, which aligns with its lowest elastic recovery of 34%. This observation indicates that plastic deformation predominates in this case.
- The high load indentation test, TEM studies and the rCLS model show that all NMMs films are predominantly plastically deformed (low flow strength $H/2.7 = 2.4$–3.0 GPa, low elastic recovery 38–54%; no radial cracks from imprint corners), plastic deformation take place mainly within the soft Cu layer, and the dislocation propagation across the incoherent interface ($fcc$-Cu vs. tetragonal-Ta) is largely excluded, respectively.

- During the deformation of Ta/Cu with $\Lambda \geq 12$ nm and due to the low flow strength of Cu and the restricted dislocation propagation across the Ta/Cu interface, the accumulated deformation interfacial energy is released by the formation of a shear band – failure of the whole structure. While in the $Ta/Cu_6$ with nanocomposite microstructure, the deformation is elasto-plastic, accompanied by the formation and propagation of dislocation, and therefore the shear bands formation is less pronounced.

### CRediT authorship contribution statement

**Daniel Karpinski:** Writing – original draft, Visualization, Validation, Methodology, Investigation, Formal analysis. **Tomas Polcar:** Writing – review & editing, Supervision, Project administration, Funding acquisition, Conceptualization. **Andrey Bondarev:** Writing – review & editing, Visualization, Validation, Methodology, Investigation, Conceptualization.

### Declaration of competing interest

The authors declare that they have no known competing financial interests or personal relationships that could have appeared to influence the work reported in this paper.

### Data availability

The data that support the findings of this study are available from the corresponding author upon reasonable request.

### Acknowledgments

CzechNanoLab project LM2018110 funded by MEYS CR is gratefully acknowledged for the financial support of the measurements at CEITEC Nano Research Infrastructure and Institute of Physics of the Czech Academy of Sciences. This work was co-funded by the European Union under the project Robotics and advanced industrial production (reg. no. CZ.02.01.01/00/22_008/0004590).

### Appendix A

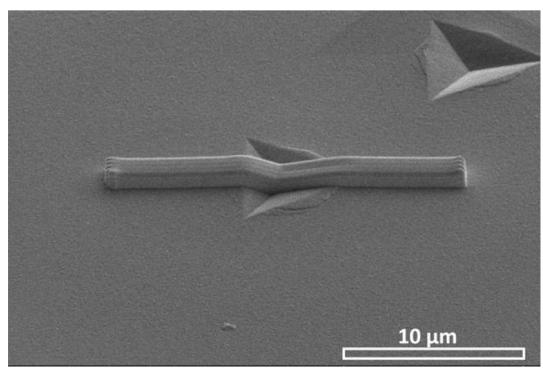

**Fig. A1.** SEM image of the imprint on a surface of $Ta/Cu_{20}$ sample with a deposited mask for further FIB milling.

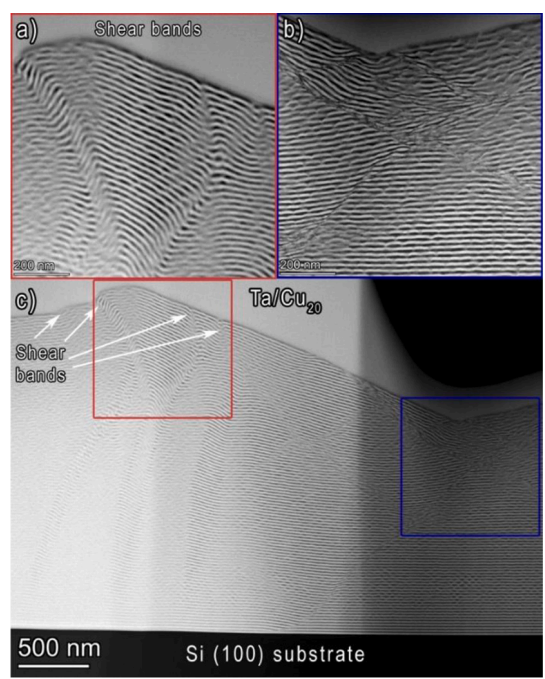

**Fig. A2.** Cross-section HAADF image of the plastically deformed $Ta/Cu_{20}$ NMM film deposited at a low $a_D$.

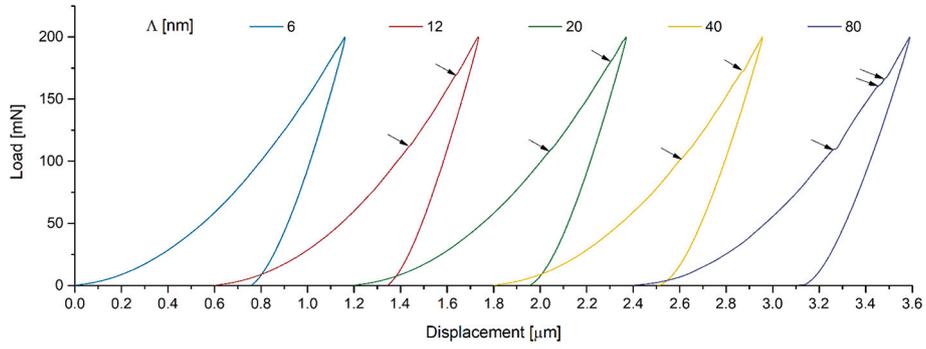

**Fig. A3.** Loading-unloading curves taken during indentations at a high load of as-deposited Ta/Cu NMMs as a function of periodicity $\Lambda$. The arrows on the loading curve represent kinks formed during the shearing of the layers.

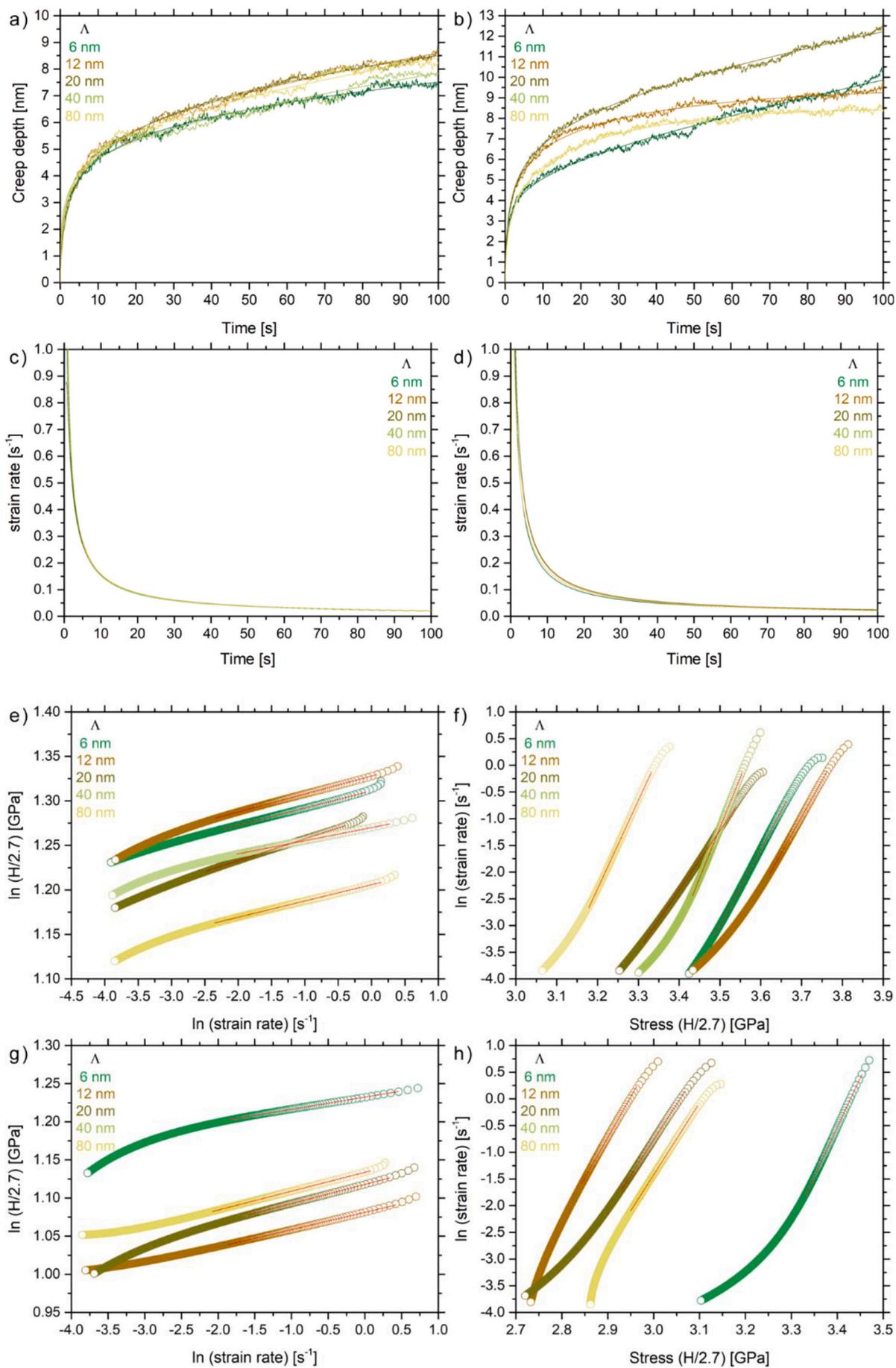

**Fig. A4.** Evaluation of the strain rate sensitivity and activation volume in the Ta—Cu NMMs deposited at a low $a_D$ (a, c, e, and g) and high $a_D$ (b, d, f and h). Depth vs. time at a maximum load (a and b). Strain rate (c and d) was determined from fitting (a and b). The strain rate sensitivity (*m*) was determined from the slope of linear fit [red lines] (e and f). The activation volume V* was determined from the slope of linear fit [red lines] (g and h). (For interpretation of the references to colour in this figure legend, the reader is referred to the web version of this article.)

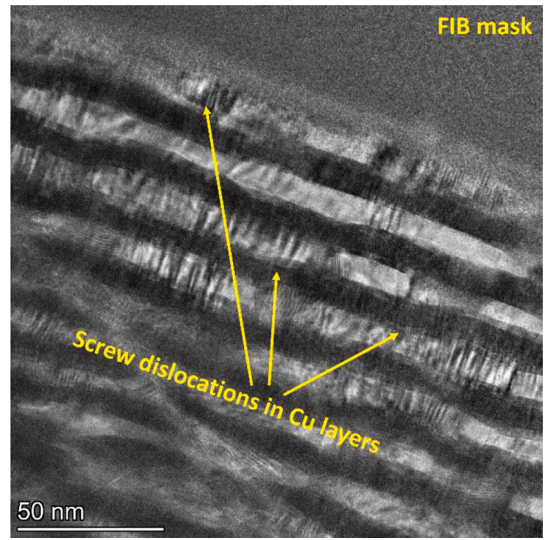

**Fig. A5.** Cross-section BF TEM image of the deformed Ta/Cu$_{20}$ NMM.